\pdfoutput=1
\documentclass[fleqn,10pt]{article}
\usepackage[utf8]{inputenc}
\usepackage[T1]{fontenc}
\usepackage[margin=1in]{geometry}
\usepackage{graphicx}
\usepackage{xcolor}
\usepackage{cite}
\usepackage{hyperref}
\hypersetup{hidelinks}
\usepackage{microtype}
\usepackage{bm}
\usepackage{siunitx}
\graphicspath{{figures/}}
\usepackage{placeins}
\usepackage{amsmath,amssymb}
\providecommand{\JournalTitle}[1]{#1}
\title{Data-center growth impacts household electricity costs more through wholesale prices than through grid reinforcement}

\author{Lingdong Zhou$^{1,\dagger}$, Wenjia Xia$^{1,\dagger}$, Jiyu Wang$^{2}$, Fei Ding$^{2,*}$, Xiangqi Zhu$^{1,*}$\\[0.5em]
\small $^{1}$School of Electrical Engineering and Computer Science, Oregon State University, Corvallis, OR, USA\\
\small $^{2}$Power Systems Engineering Center, National Laboratory of the Rockies, Golden, CO, USA\\
\small $^{\dagger}$These authors contributed equally to this work.\\
\small $^{*}$Correspondence: Fei Ding (\href{mailto:Fei.Ding@nlr.gov}{Fei.Ding@nlr.gov}) and Xiangqi Zhu (\href{mailto:xiangqi.zhu@oregonstate.edu}{xiangqi.zhu@oregonstate.edu})}
\date{}

\begin{document}
\pagestyle{plain}
\maketitle

\begin{abstract}
Rapid data-center expansion is raising concerns about electricity affordability, yet how the costs of serving these large new loads reach residential customers remains poorly understood. Here, we quantify the potential impact of 371 announced Texas data-center projects, totaling 38.9~GW by 2030, on residential electricity costs by linking power system expansion, wholesale electricity markets, transmission reinforcement, and retail cost allocation. We find that accommodating this demand increases modeled residential electricity costs in one representative case by 1.64~cents/kWh, with 68\% arising from higher wholesale electricity prices and 32\% from regulated transmission charges.  Allowing new gas generation near data centers reduces the 2023 residential cost increase to 0.92~cents/kWh, whereas excluding data centers from the transmission cost-sharing base raises it to 1.87~cents/kWh. These findings reveal that the residential cost burden of data-center growth is driven primarily by wholesale electricity price increases, rather than direct transmission investment costs, and depends critically on generation siting, infrastructure planning, and cost allocation.
\end{abstract}

\section*{Introduction}
For much of the past two decades, growth in digital services did not translate into comparable growth in data-center electricity demand because efficiency improvements in computing hardware, cooling, and facility operations offset much of the increase in computing activity\cite{Koomey2008,Masanet2020}. The rapid expansion of artificial intelligence and other compute-intensive workloads is now breaking this pattern\cite{Patterson2022,IEA2025}. After years of comparatively flat increase in demand, utilities and reliability planners in the United States project significant load growth driven substantially by data centers\cite{NERC2024,Shehabi2024}. Proposed facilities can require hundreds of megawatts to several gigawatts of capacity, operate at high utilization rates, develop on compressed timelines, and cluster geographically\cite{Rand2024,ERCOT2025}. Their growth therefore raises an economic question: how are the resulting system costs incurred, through which market and regulatory channels are they recovered, and who ultimately bears them\cite{Martin2025}? These effects are already emerging in electricity markets. Data-center load growth was the dominant driver of the record increase in PJM capacity prices for 2025/26\cite{IEEFA2025}, while the U.S. Energy Information Administration projects that, under a high-demand scenario, data-center growth could increase wholesale electricity prices in the Electric Reliability Council of Texas (ERCOT) by nearly 80\% by 2027\cite{EIA2026}.

Most assessments of data-center electricity demand focus on aggregate consumption, environmental impacts, or resource adequacy\cite{Siddik2021,Guidi2024,ChenX2025}. Less is known about how the costs created by large new loads propagate through the electricity system to households. That propagation is spatial. The same quantity of new demand can be relatively straightforward to accommodate near strong generation and transmission, yet difficult to serve behind a constrained corridor. In the latter case, generation added elsewhere reaches the load only up to the capacity of the corridor.

The downstream consequences also pass through distinct economic channels. New load changes dispatch, congestion, and locational marginal prices (LMPs)\cite{Bohn1984,Schweppe1988,Hogan1992}; it changes the economically preferred mix and location of generation, storage, and transmission expansion; and transmission investment is recovered through regulated network charges outside the wholesale market\cite{Joskow2005,Lyons2000}. Wholesale and regulated costs then reach retail providers, utilities, and, depending on market design and rate regulation, customers\cite{Borenstein2022}. Quantifying household impacts therefore requires an integrated assessment of siting, physical deliverability, infrastructure expansion, wholesale repricing, regulated cost allocation, and retail exposure.

Texas provides a revealing setting in which to trace this chain. ERCOT combines rapid load growth\cite{ERCOT2025}, regions constrained by transmission capacity, a wholesale market in which prices can rise steeply when supply is scarce\cite{Cramton2017,Hogan2013,Potomac2024}, and active retail competition\cite{Hortacsu2017,Zarnikau2023}. The February 2021 winter blackout also demonstrated how quickly physical stress can become financial stress for providers and customers\cite{Busby2021,King2021,Littlechild2021}. Together with the scale and uneven geography of announced data-center development, these characteristics allow us to examine how a large-load scenario propagates from interconnection to the residential bill.

Here we construct an announced-development scenario comprising 371 Texas data-center projects, each assigned to its reported location, totaling \SI{38.9}{\giga\watt} by 2030 and operating at high utilization. We follow its consequences through three linked levels of analysis: system planning and wholesale prices, engineering reinforcement needed for secure network operation, and the wholesale and regulated cost channels that reach residential customers. To account for year-to-year variations in electricity demand and renewable generation, we evaluate the same 2030 data-center scenario using historical demand and renewable-availability profiles from five weather years (2019--2023). Each simulation covers May through October, focusing on the high-demand summer season in Texas which determines peak-based transmission cost allocation.

The analysis identifies a common mechanism across these layers: the consequences of large-load growth depend on whether dispatchable generation can be built where the load connects; when it cannot, deliverability limits what can be served. When new gas capacity can expand only at existing plant sites, generation and storage alone supply more than 99.95\% of total demand but leave substantial modeled scarcity exposure behind constrained corridors, whereas coordinated generation and transmission expansion reduces annualized investment by 73\% and confines that exposure to a few hours per season. The resulting modeled peak-season residential cost increase is 1.64~cents/kWh, with most of the increase arising through wholesale repricing rather than the regulated transmission reinforcement charge.

\section*{Results}

\subsection*{Wholesale prices carry most of the residential electricity cost impact}

\begin{figure}[!htbp]
\centering
\includegraphics[width=\linewidth]{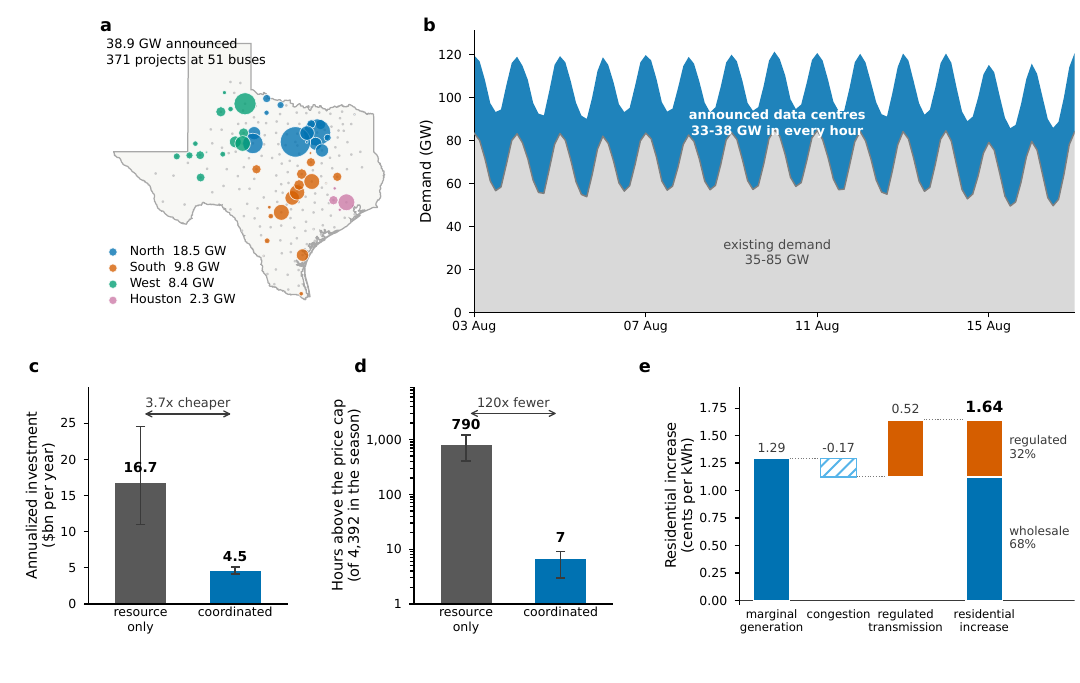}
\caption{\textbf{A nearly constant \SI{38.9}{\giga\watt} of announced demand, the two ways of serving it, and what reaches the residential bill.}
\textbf{a}~The announced 2030 scenario: 371 projects aggregated to 51 buses of the ERCOT model, totaling \SI{38.9}{\giga\watt}. Bubble area is proportional to the capacity at each bus; colors give the ERCOT zone, with zonal totals in the legend; grey dots mark the remaining model buses.
\textbf{b}~Existing ERCOT demand (grey) with the announced data center load stacked on top (blue), over the two weeks around the 2023 summer peak. Existing demand swings between 35 and \SI{85}{\giga\watt}; the data centers add 33--\SI{38}{\giga\watt} in every hour of the season.
\textbf{c}~Annualized investment needed to serve that load when only generation and storage may expand, and when transmission may expand alongside them. Bars give the mean across the five weather years 2019--2023, whiskers the range.
\textbf{d}~Hours per season, of 4,392 (each three-hour interval counts three hours), in which at least one bus prices above the \$5,000/MWh offer cap, for the same two pathways (logarithmic scale).
\textbf{e}~The 1.64~cents/kWh residential increase during the peak season in 2023, resolved into the wholesale channel (marginal generation and congestion, 68\%) and the regulated transmission charge (32\%); the congestion term is negative in 2023 and is drawn as a hatched step. Panels c--e are developed in full in Figs.~\ref{fig:all2030}, \ref{fig:attribution} and \ref{fig:customer_rep}.}
\label{fig:framework}
\end{figure}

The announced \SI{38.9}{\giga\watt} data-center scenario adds a large, geographically concentrated, and nearly constant block of demand to ERCOT. The 371 projects are concentrated at 51 buses, with aggregate data-center demand remaining at roughly 33--\SI{38}{\giga\watt} throughout the peak season even as existing system demand varies much more strongly (Fig.~\ref{fig:framework}). Serving this load therefore requires not only sufficient generation, but also enough network capacity to deliver power to where the new demand connects.

How the system expands substantially changes the cost of meeting that demand. We compare two pathways: generation and storage alone (``resource only'' or ``generation + storage'' in the figures) and coordinated expansion, in which transmission may expand alongside them (``full expansion'' in the figures). Generation and storage alone supply more than 99.95\% of the load, but leave persistent localized scarcity, with at least one bus above the \$5{,}000/MWh market offer cap in 136--407 three-hour intervals per season. Allowing transmission to expand alongside generation reduces mean annualized investment from \$16.7~billion to \$4.54~billion, a 73\% reduction, while confining above-cap exposure to one to three intervals per season.
With gas limited to existing plant sites, the cost of serving the load is therefore set mainly by delivering power across constrained corridors.

These system effects ultimately appear in residential electricity costs. Although substantial transmission reinforcement is required to accommodate the new load, most of the household impact comes through the wholesale market rather than through direct recovery of transmission investment. In the 2023 weather year case, the peak-season residential increase is 1.64~cents/kWh, of which 1.124~cents/kWh, or 68\%, arises through wholesale repricing and 0.520~cents/kWh, or 32\%, through regulated transmission charges.

The remainder of the Results section traces this finding from its physical origin, where data centers connect and whether power can reach them, through infrastructure expansion with and without gas plants beside the data centers, and through wholesale price formation, to the costs ultimately reaching retailers and households.

\subsection*{Where data centers connect determines whether they can be served}

\begin{figure}[!htbp]
\centering
\includegraphics[width=\linewidth]{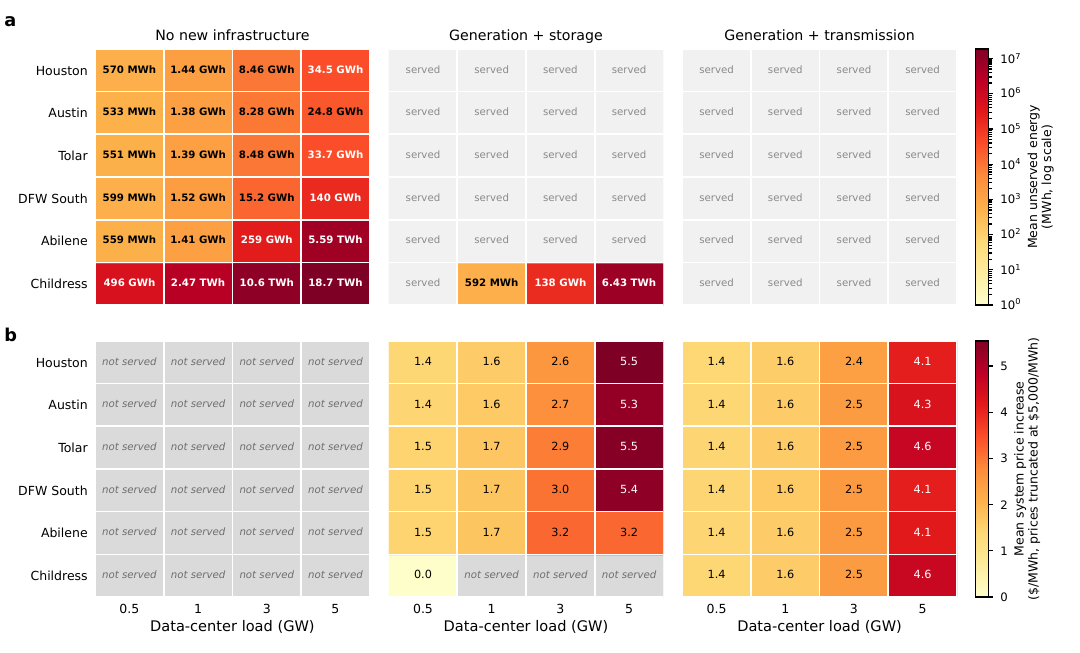}
\caption{\textbf{Where the load connects, rather than how large it is, strongly shapes whether it can be served when dispatchable generation cannot be built on site.} Six representative locations, four load scales, and three system responses; every cell is a mean over the five weather years 2019--2023. The two panels share the same grid and are read column by column.
\textbf{a}~Unserved energy on a logarithmic color scale. Cells below \SI{1}{\mega\watt\hour} per season fall within solver tolerance and are labelled \textsc{served}. Severity spans three orders of magnitude: at \SI{5}{\giga\watt} with no new infrastructure, Houston leaves \SI{34.5}{\giga\watt\hour} unserved and Childress \SI{18.7}{\tera\watt\hour}. Generation and storage clear every location except Childress; adding transmission clears all of them.
\textbf{b}~Increase in the average system price relative to the baseline without data centers, with prices truncated at the \$5,000/MWh offer cap (\$/MWh), shown only where the load is fully served in all five weather years; no case without new infrastructure qualifies. Cells with residual unserved load are greyed because their prices are set by the cost assigned to unserved load rather than by market clearing.}
\label{fig:siting}
\end{figure}

The location of data centers can substantially change the grid's ability to accommodate new electricity demand. To isolate this effect, we place the same amount of additional demand, ranging from 0.5 to 5~GW, at six representative locations across ERCOT and compare the outcomes under different infrastructure expansion pathways (Fig.~\ref{fig:siting}). Without new infrastructure, adding 5~GW leaves 25--140~GWh of electricity demand unserved per season at Houston, Austin, Tolar, and southern Dallas--Fort Worth, compared with 5.6~TWh at Abilene and 18.7~TWh at Childress (Fig.~\ref{fig:siting}a). Even a 0.5~GW addition at Childress leaves 496~GWh unserved. Thus, identical data-center demand can produce differences of nearly three orders of magnitude in unserved energy depending on where it connects.

Infrastructure expansion can alleviate these constraints, but its effectiveness also depends on location. When new gas generation is restricted to existing plant sites, expanding generation and storage eliminates substantial unserved energy at five of the six locations across all tested demand levels. Childress remains an exception, with 6.43~TWh unserved at 5~GW despite substantial investment. Allowing transmission to expand alongside generation and storage serves all tested demand at every location. Alternatively, permitting new gas generation directly at the data-center locations also serves all six 5~GW cases without transmission expansion in the 2023 weather year. At Childress, this option costs \$0.64~billion annually, compared with \$0.73~billion for coordinated expansion and \$3.15~billion for generation and storage alone, which still leave demand unserved. These comparisons show that transmission expansion and local dispatchable generation can provide alternative ways to overcome location-specific delivery constraints.

The infrastructure investment required to accommodate data-center demand also varies by location. Under coordinated expansion, annualized investment for 5~GW ranges from \$0.137~billion at Austin to \$0.252~billion at Childress, averaged across the five weather years. Where generation and storage alone can fully serve the load, their investment requirements are 1.2 times those of coordinated expansion at Austin, 1.3 times at Houston, and 2.9 times at Abilene. Despite these differences in infrastructure requirements, the increase in average system-wide wholesale electricity prices under coordinated expansion is relatively similar across locations, ranging from \$4.1 to \$4.6/MWh for the same 5~GW of additional demand (Fig.~\ref{fig:siting}b). These findings show that data-center location strongly influences electricity deliverability and infrastructure requirements, even when its impact on average system-wide wholesale prices is similar.

\subsection*{Coordinated expansion meets the announced load at lower cost}

\begin{figure}[!htbp]
\centering
\begin{minipage}[t]{0.48\linewidth}
{\sffamily\bfseries a}\\[0.1em]
\includegraphics[width=\linewidth]{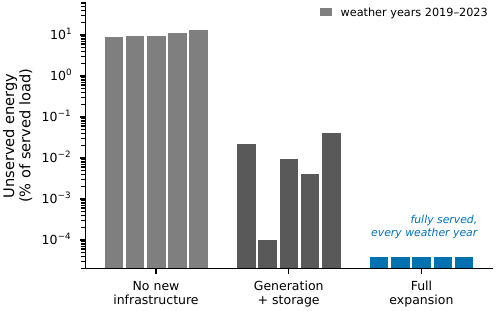}
\end{minipage}\hfill
\begin{minipage}[t]{0.48\linewidth}
{\sffamily\bfseries b}\\[0.1em]
\includegraphics[width=\linewidth]{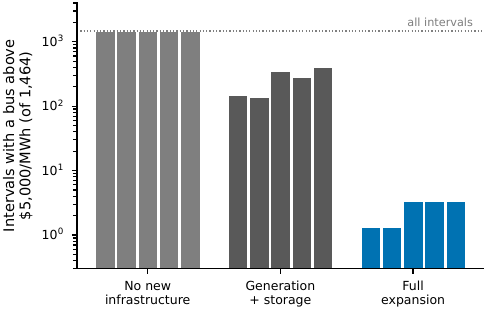}
\end{minipage}\\[0.9em]
\begin{minipage}[t]{0.48\linewidth}
{\sffamily\bfseries c}\\[0.1em]
\includegraphics[width=\linewidth]{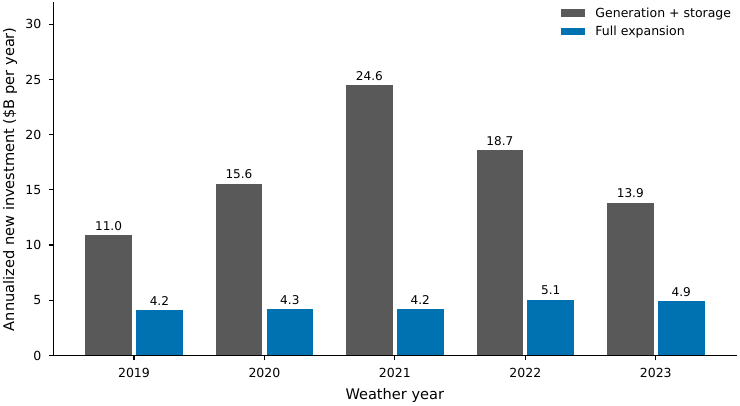}
\end{minipage}\hfill
\begin{minipage}[t]{0.48\linewidth}
{\sffamily\bfseries d}\\[0.1em]
\includegraphics[width=\linewidth]{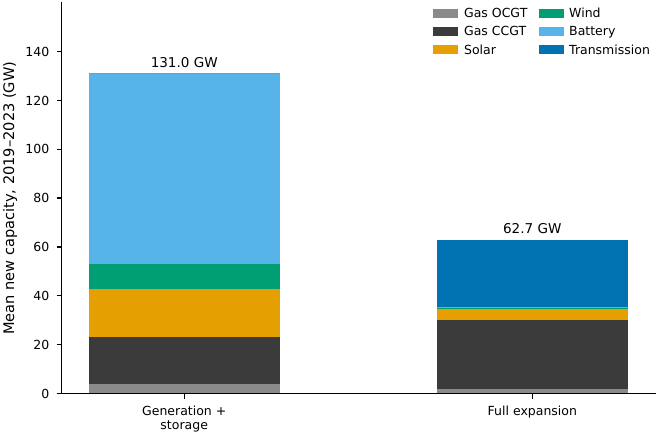}
\end{minipage}
\caption{\textbf{Coordinated generation and transmission expansion serves the announced load at lower cost and with little residual scarcity.}
\textbf{a}~Unserved energy as a percentage of served load, by system response and weather year (logarithmic scale; one bar per weather year). With no new infrastructure, this ratio is 9.5--13.4\%. Generation and storage reduce this to 0.0001--0.042\%, and coordinated expansion serves the load in full.
\textbf{b}~Number of three-hour intervals, of 1,464 in the season, containing at least one bus priced above the \$5,000/MWh offer cap. Generation and storage leave 136--407 such intervals; coordinated expansion leaves one to three.
\textbf{c}~Annualized new investment by weather year for the generation-and-storage case (\$11.0--24.6~billion per year) and the coordinated expansion case (\$4.2--5.1~billion per year).
\textbf{d}~Mean new capacity across the five weather years, by technology. Without transmission expansion the optimization builds approximately \SI{77.9}{\giga\watt} of battery power on average; with transmission expansion the build shifts to generation and transmission with negligible storage.}
\label{fig:all2030}
\end{figure}

The location experiments above show that transmission constraints can determine whether large data-center loads can be served affordably. We next scale this delivery challenge to the full \SI{38.9}{\giga\watt} announced scenario and evaluate it across five weather years, 2019--2023, to account for year-to-year variation in electricity demand and renewable availability. Without new infrastructure, unserved energy equals 9.5--13.4\% of served load across these years, and at least one bus exceeds the \$5{,}000/MWh offer cap in every three-hour interval of the May--October season (Fig.~\ref{fig:all2030}a,b).

Adding generation and storage substantially reduces the electricity shortage but does not fully resolve the transmission constraints. This pathway serves more than 99.95\% of the load, yet localized scarcity persists, with 136--407 three-hour intervals per season in which at least one bus exceeds the market offer cap (Fig.~\ref{fig:all2030}a,b). Accommodating the demand also requires substantial investment, ranging from \$11.0 to \$24.6~billion per year across the five weather years, with an average of \$16.7~billion (Fig.~\ref{fig:all2030}c). The resulting expansion includes approximately 77.9~GW of battery storage power capacity on average (Fig.~\ref{fig:all2030}d). These results indicate that additional generation and storage alone cannot efficiently overcome the constraints on delivering electricity to geographically concentrated data-center loads.

Allowing transmission to expand alongside generation and storage substantially improves both cost and reliability outcomes. Coordinated expansion serves all demand across the five weather years and reduces above-cap exposure to just one to three three-hour intervals per season. Annualized investment falls to \$4.2--5.1~billion per year, averaging \$4.54~billion, a 73\% reduction compared with generation and storage expansion alone (Fig.~\ref{fig:all2030}c). The resulting portfolio includes an average of 35.5~GW of new generation and 27.2~GW of transmission capacity, with only 0.05~GW of storage (Fig.~\ref{fig:all2030}d). These findings demonstrate the importance of coordinated transmission expansion in serving large, geographically concentrated loads when new gas generation is restricted to existing plant sites. We next examine how these outcomes change when gas generation can also be built near the data centers.

\subsection*{Where new gas plants can be built shapes household electricity costs}

The preceding results show that transmission expansion substantially reduces the cost of accommodating data-center demand when new gas generation is restricted to existing power plant sites. However, new gas plants can also be developed at other locations, potentially closer to large electricity consumers. To understand how generation siting affects infrastructure investment and residential electricity costs, we compare two scenarios. The first allows new gas capacity only at buses with existing gas generation, representing expansion at established generation sites where electricity and fuel infrastructure may already be available. The second also allows new gas plants at all 51 data-center buses, representing a more flexible siting scenario in which generation can be built closer to the new demand. This comparison helps determine how much of the transmission investment is driven by restrictions on generation locations.

\begin{figure}[!htbp]
\centering
\includegraphics[width=\linewidth]{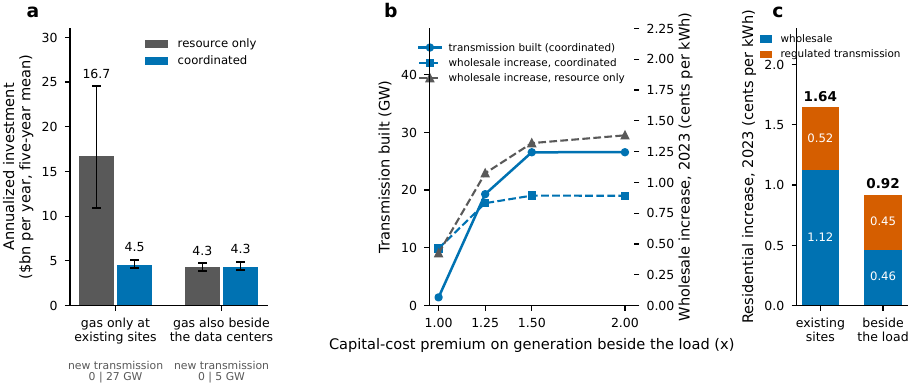}
\caption{\textbf{Whether gas generation can be built beside the data centers decides how much transmission the announced load needs, and lowers what households pay.} \textbf{a}~Annualized investment to serve the announced load when generation and storage alone may expand (grey) and when transmission may expand alongside them (blue), with new gas capacity allowed only at existing plant sites or also at the 51 data-center buses. Bars give the mean over the weather years 2019--2023, whiskers the range; labels give the mean transmission built. \textbf{b}~Capital-cost premium on the local gas candidates (2023): transmission built under coordinated expansion (solid, left axis) and the wholesale increase under both pathways (dashed, right axis). \textbf{c}~Residential increase in 2023 under the two candidate sets, split into the wholesale channel and the regulated transmission charge; both charges come from the same Texas2k screening.}
\label{fig:regimes}
\end{figure}

Allowing gas generation near data centers substantially reduces the need for transmission expansion (Fig.~\ref{fig:regimes}a). When gas generation is restricted to existing plant sites, generation and storage expansion alone costs an average of \$16.7~billion annually, compared with \$4.54~billion for coordinated expansion. When gas generation is also allowed at data-center locations, the two pathways converge to approximately \$4.3~billion annually, with investment differences of only 0.1--5.4\% across the five weather years. Under coordinated expansion, average transmission additions fall from 27.2 to 5.4~GW, while no battery storage is needed. Both pathways also reduce above-cap price exposure to one to three three-hour intervals per season. In the 2023 weather year, new combined-cycle gas generation is built at 44 of the 51 data-center buses, totaling 26.9~GW. These findings indicate that building dispatchable generation closer to data-center demand can reduce the need to transport electricity over long distances, substantially lowering infrastructure requirements.

The benefits also extend to residential electricity costs (Fig.~\ref{fig:regimes}c). In the 2023 weather year, allowing gas generation at data-center locations reduces the modeled residential cost increase from 1.64 to 0.92~cents/kWh. Most of this reduction comes from lower wholesale electricity prices, whose contribution falls from 1.124 to 0.463~cents/kWh. The regulated transmission charge decreases more modestly, from 0.520 to 0.453~cents/kWh. Although local generation substantially reduces transmission expansion in the planning model, branch-level engineering screening still identifies considerable reinforcement needs to accommodate the full data-center load under single-branch outages. The corresponding estimated transmission reinforcement investment decreases from \$32.42 to \$28.26~billion. Thus, locating generation closer to data centers can substantially reduce the residential cost impact, primarily through lower wholesale electricity prices.

The economic attractiveness of local gas generation nevertheless depends on its construction cost. When the capital cost of new gas plants at data-center locations increases by 25\%, coordinated expansion again favors more transmission, with additions rising to 19.3~GW in the 2023 weather year (Fig.~\ref{fig:regimes}b). At a 50\% premium, transmission additions reach 26.5~GW. These results show that the preferred balance between local generation and transmission expansion depends on the relative costs of the two options. Additional sensitivity analyses of residential electricity costs under different weather years, fuel prices, scarcity assumptions, data-center development levels, and cost-allocation parameters are summarized in Table~\ref{tab:ranges}.

\begin{table}[!htbp]
\centering
\caption{\textbf{Range of the modeled residential increase.} Cents per kWh; 2023, gas only at existing plant sites, and the total of the flat-load wholesale component and the residential regulated charge unless stated.}
\label{tab:ranges}
\begin{tabular}{p{11.2cm}c}
\hline
Quantity & Range \\
\hline
Residential increase by weather year, 2019--2023 & 1.28--2.26 \\
Residential increase with gas generation beside the load & 0.92 \\
Gas price $\times$0.5 / $\times$1.5 & 1.11 / 1.56 \\
Residential peak ratio $\phi$ 1.8--2.5 & 1.59--1.77 \\
Value of lost load $10^{4}$--$5\times10^{4}$~\$/MWh & 1.42--2.08 \\
Wholesale component under ERCOT's reserve-price rule, 2023 / 2019--2022 & 0.81 / 2.34--2.89 \\
Wholesale component weighted by the demand of existing customers & 1.53 \\
Wholesale component at 25 / 50 / 75\% of the announced load (1.12 at 100\%) & 0.67 / 0.71 / 0.76 \\
Wholesale component without already-operating capacity (\SI{30.9}{\giga\watt}) & 0.77 \\
Regulated transmission charge, all weather years and the 2023 local-gas portfolios, main allocation rule & 0.44--0.65 \\
\hline
\end{tabular}
\end{table}

\begin{figure}[!htbp]
\centering
\includegraphics[width=\linewidth]{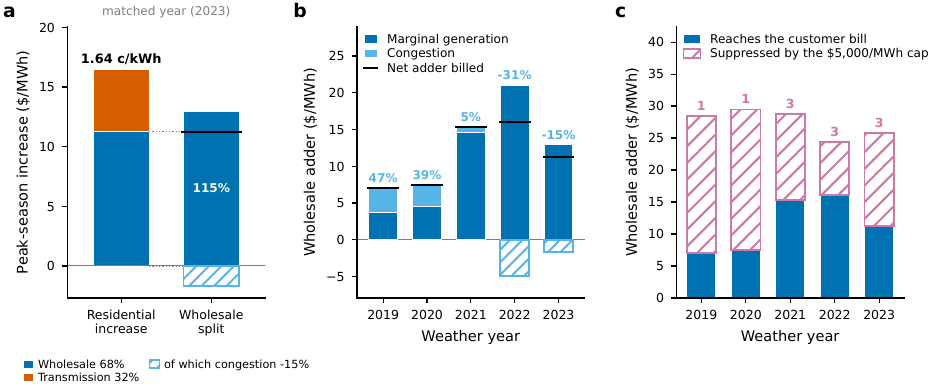}
\caption{\textbf{Attribution of the residential cost increase and the wholesale price effect.} Each term is reported at the level on which it is defined, and the two levels have different denominators.
\textbf{a}~Composition of the residential increase in 2023 (\$16.44/MWh, 1.64~cents/kWh, for a residential customer whose demand at the four system peaks is twice its average). Left bar: wholesale (68.4\%) against regulated transmission delivery (31.6\%). Right bar: the wholesale component resolved into marginal-generation and congestion terms; the congestion term is negative and is drawn below the axis.
\textbf{b}~The wholesale increase that reaches the bill, by weather year, split into marginal generation and congestion; figures give the congestion share. The annual congestion share ranges from $+47\%$ to $-31\%$; averaged in \$/MWh across the five weather years, the congestion contribution is close to zero. Negative values are drawn below the axis, and the black tick marks the net billed adder.
\textbf{c}~Wholesale increase that reaches the customer bill (solid) against the above-cap value suppressed by the \$5,000/MWh offer cap (hatched). Figures give the number of three-hour intervals, of 1,464, that contain a bus above the cap. The above-cap component can exceed the billed increase while arising in 3--9 hours per season.}
\label{fig:attribution}
\end{figure}

\begin{figure}[!htbp]
\centering
\includegraphics[width=\linewidth]{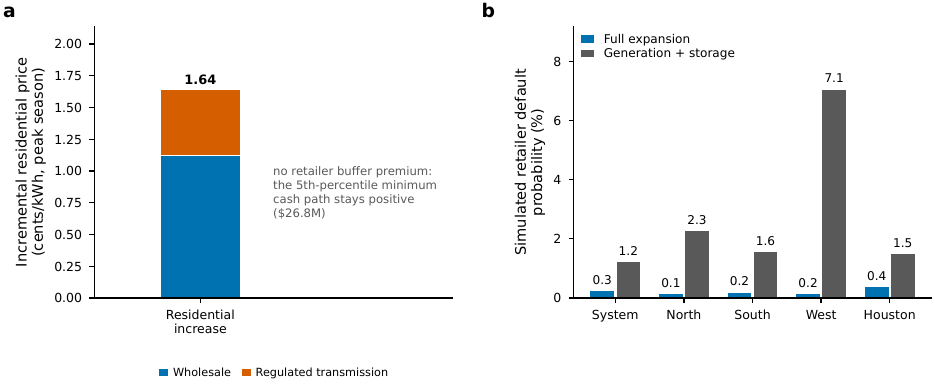}
\caption{\textbf{Customer-price channels in the 2023 weather year and retail-provider financial exposure for the announced 2030 scenario.}
\textbf{a}~Increase in customer price, split into the wholesale component (1.124~cents/kWh) and the regulated transmission charge (0.520~cents/kWh for a residential customer whose demand at the four system peaks is twice its average); no retail-provider liquidity premium arises under the 95\% protection criterion because the fifth percentile of minimum cash remains positive.
\textbf{b}~Simulated retail-provider default probability by pricing zone under repriced procurement (5,000 Monte Carlo paths each), comparing coordinated generation-and-transmission expansion with generation-and-storage expansion. The generation-and-storage pathway raises system default probability from 0.3\% to 1.2\%, and from 0.2\% to 7.1\% in West.}
\label{fig:customer_rep}
\end{figure}

\subsection*{Planning determines the costs; allocation determines who pays}

The costs of accommodating data-center growth reach residential customers through two main channels: higher wholesale electricity prices and regulated charges for transmission reinforcement. Under coordinated generation and transmission expansion, the average wholesale electricity price in the 2023 weather year increases from \$26.43/MWh to \$37.67/MWh, adding 1.124~cents/kWh to residential electricity costs. Separately, the transmission reinforcement required to accommodate the new demand involves \$32.42~billion in initial investment, equivalent to \$2.719~billion annually. Allocating this cost based on customers' contributions to system peak demand adds another 0.520~cents/kWh for a representative residential customer whose peak demand is twice its average load. Together, these two channels produce a modeled residential cost increase of 1.64~cents/kWh, with 68\% arising from wholesale price increases and 32\% from regulated transmission charges (Fig.~\ref{fig:attribution}a).

The wholesale price increase is driven primarily by the higher marginal cost of supplying additional electricity, although transmission congestion also contributes in some years. Across the five weather years, the wholesale component ranges from 0.71 to 1.61~cents/kWh (Fig.~\ref{fig:attribution}b). Congestion accounts for 39--47\% of the increase in 2019--2020 and 5\% in 2021, but contributes negatively in 2022--2023, when the expanded network experiences less congestion than the baseline system without data centers. Averaged across the five years, the congestion contribution is close to zero. The wholesale cost increase therefore arises mainly from the additional generation needed to serve data-center demand. The estimates also depend on how scarcity is priced. Under coordinated expansion, modeled prices exceed the \$5{,}000/MWh market offer cap at some buses during one to three three-hour intervals per season. The additional scarcity value above the cap averages \$15.95/MWh across the five weather years but is excluded from the main residential cost estimate (Fig.~\ref{fig:attribution}c). When ERCOT's reserve-pricing mechanism is represented, the wholesale component is 0.81~cents/kWh in 2023 and 2.34--2.89~cents/kWh in 2019--2022.

Higher wholesale electricity costs do not necessarily translate immediately into higher residential bills. Under fixed-price retail contracts, electricity retailers initially absorb changes in procurement costs and regulated transmission charges until these costs can be reflected in future retail prices. To examine this exposure, we simulate a retail electricity provider serving a constant 1~GW load. Under coordinated expansion, repricing electricity procurement to account for data-center demand increases the retailer's average procurement cost from \$36.5/MWh to \$48.1/MWh and reduces its average margin from \$42.0/MWh to \$27.5/MWh. The simulated probability of exhausting its cash reserves increases from 0.04\% to 0.26\% (Fig.~\ref{fig:customer_rep}a). However, the retailer retains sufficient liquidity under the 95\% protection criterion, so no additional liquidity premium is included in the residential cost estimate.

The choice of infrastructure expansion also affects retailers' financial exposure. With gas generation restricted to existing plant sites, relying on generation and storage without transmission expansion increases the simulated probability of exhausting cash reserves from 0.26\% under coordinated expansion to 1.24\% system-wide, and from 0.16\% to 7.06\% in the West zone (Fig.~\ref{fig:customer_rep}b). Although coordinated expansion requires regulated transmission investment, it reduces wholesale procurement costs by \$7.35/MWh, exceeding the \$2.6--3.3/MWh regulated transmission charge faced by the modeled flat-load retailer. This comparison shows that transmission investment can reduce retailers' financial exposure by lowering wholesale procurement costs, even after accounting for its regulated cost recovery.

\begin{figure}[!htbp]
\centering
\includegraphics[width=\linewidth]{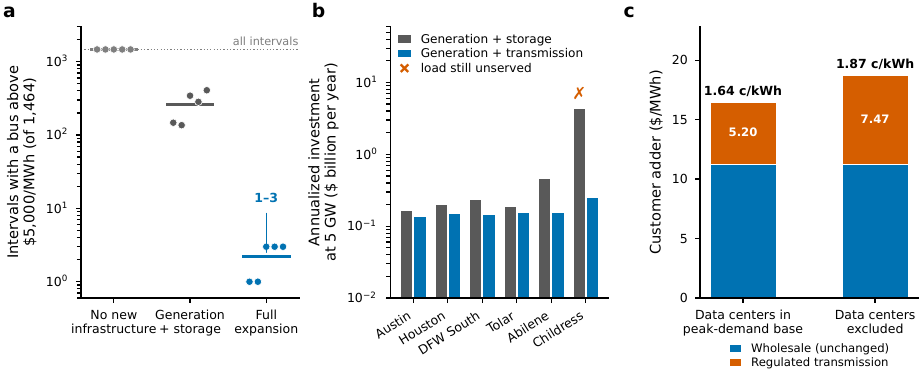}
\caption{\textbf{Different levers act on different stages of the pathway from data-center load to customer cost.}
\textbf{a}~Data-center flexibility targets residual scarcity. Points give the number of three-hour intervals, of 1,464, containing a bus above \$5,000/MWh in each weather year; bars give the five-year mean. Coordinated expansion leaves one to three such intervals, generation and storage alone 136--407, and the existing system all 1,464.
\textbf{b}~Siting changes the infrastructure required for delivery. Annualized investment to serve \SI{5}{\giga\watt} at each representative location, with and without transmission expansion (logarithmic scale). Crosses mark cases in which load remains unserved after the investment shown.
\textbf{c}~Cost allocation changes only the regulated component. Excluding data centers from the system-peak demand base over which transmission cost is spread raises the representative residential total from 1.64 to 1.87~cents/kWh while leaving the wholesale component unchanged.}
\label{fig:levers}
\end{figure}

Beyond infrastructure planning, how transmission costs are allocated also determines the burden borne by residential customers. Under the main allocation assumption, data centers contribute approximately \$0.82~billion toward the \$2.719~billion annual transmission revenue requirement. If data-center demand is excluded from the system peak-demand base used to allocate these costs, the representative residential transmission charge increases from 0.520 to 0.747~cents/kWh, raising the total residential cost increase from 1.64 to 1.87~cents/kWh (Fig.~\ref{fig:levers}c). This change leaves wholesale electricity prices and infrastructure investment unchanged but shifts a larger share of transmission costs to existing customers. These findings show that infrastructure planning and cost-allocation rules influence residential electricity costs through distinct mechanisms: planning affects the costs incurred, while allocation determines how those costs are distributed.

\FloatBarrier

\section*{Discussion}

Our findings show that the residential electricity cost impacts of data-center growth extend beyond the infrastructure built to accommodate new demand. In the 2023 weather year, wholesale electricity price increases account for 68\% of the modeled residential cost increase, with regulated transmission charges contributing the remaining 32\%. The wholesale effect arises because additional data-center demand changes electricity prices for existing consumers across the system. Thus, even when data centers contribute to transmission investment, their demand can still increase electricity costs for households. This distinction highlights the importance of considering both wholesale market responses and regulated infrastructure cost recovery when assessing electricity affordability.

Our results point to three complementary approaches for managing these impacts: data-center flexibility, siting and infrastructure planning, and transmission cost allocation (Fig.~\ref{fig:levers}). Each acts on a different part of the pathway from new electricity demand to residential costs. Flexibility can reduce demand during periods of system stress, siting and infrastructure planning determine how efficiently new demand can be served, and cost allocation determines how regulated infrastructure costs are shared.

Data-center flexibility could help alleviate the remaining scarcity after infrastructure expansion. Under coordinated expansion, uncapped modeled electricity prices exceed the \$5{,}000/MWh market offer cap at some buses during only one to three three-hour intervals per season (Fig.~\ref{fig:levers}a). These periods provide potential targets for demand response, workload scheduling, and temporal or spatial load shifting\cite{Wierman2014,Radovanovic2023,Norris2025}. However, flexibility is not explicitly modeled in this study, and its benefits depend on the amount and duration of demand that can be shifted. These results identify when flexibility could be particularly valuable, but do not establish how much it would reduce residential electricity costs.

Siting and infrastructure planning address the physical constraints underlying these costs. Figure~\ref{fig:levers}b shows that accommodating the same 5~GW of data-center demand can require substantially different infrastructure investments depending on where it connects. Generation siting also matters. When new gas generation is restricted to existing plant sites, coordinated transmission expansion substantially reduces the investment needed to serve data centers. Allowing generation near the new loads further reduces transmission requirements and lowers the modeled residential cost increase, primarily through lower wholesale prices. These findings extend previous research on coordinated infrastructure expansion\cite{Brown2018,Manocha2025,MoradiSepahvand2021,Victoria2020,Lee2026} by connecting infrastructure choices to residential electricity costs. However, local generation and transmission are not always interchangeable. Their relative attractiveness depends on construction costs, land availability, fuel supply, permitting, and other site-specific constraints. The results therefore support evaluating generation and transmission options together when planning for large new loads.

Transmission cost allocation addresses a different question: how the costs of infrastructure investment are distributed. In our analysis, excluding data centers from the system-peak demand base used to allocate transmission costs increases the modeled residential cost impact from 1.64 to 1.87~cents/kWh, without changing infrastructure investment or wholesale electricity prices (Fig.~\ref{fig:levers}c). Cost allocation therefore changes the burden on existing customers without reducing the underlying cost of serving new demand. This distinction is particularly relevant in Texas, where Senate Bill 6 of 2025 requires a review of the four-coincident-peak transmission cost-allocation method\cite{TXSB6}. Together, the three approaches highlight the need to consider both how electricity system costs are incurred and how they are distributed.

Retail electricity contracts introduce an additional consideration by affecting when these costs reach households. Under fixed-price contracts, retailers may initially absorb higher wholesale procurement costs and transmission charges, reducing their margins and increasing financial exposure. Our simulations show that coordinated transmission expansion reduces this exposure because the associated savings in wholesale procurement costs exceed the additional regulated transmission charges. Although the modeled retailer requires no additional liquidity buffer under the adopted 95\% protection criterion, financial risks could create further costs if retailers need additional reserves and recover them through future electricity prices. The timing of household bill increases therefore depends partly on retail contracting and procurement arrangements.

Several limitations should be considered when interpreting these findings. First, the analysis uses announced data-center projects and historical operating conditions from May to October, so the results depend on future project realization and do not capture winter conditions. Second, wholesale price estimates depend on the temporal resolution, perfect-foresight assumptions, and scarcity-pricing rules of the market model. Third, transmission costs are estimated using a synthetic network and a single-pass reinforcement procedure that leaves some violations unresolved. The wholesale and transmission models are also coupled sequentially, without recalculating prices on the reinforced network. Consequently, the reported residential cost increases represent modeled impacts under the specified assumptions, not forecasts of actual household electricity bills.

\section*{Methods}

\subsection*{Power system planning and wholesale price model}

\subsubsection*{ERCOT network, dispatch, and LMP calculation}

We use PyPSA-USA\cite{Tehranchi2025}, built on the PyPSA framework\cite{PyPSA,Horsch2018,Schlachtberger2017}, to jointly optimize capacity expansion and dispatch over a period of six months covering peak demand. The model minimizes annualized investment and operating costs,
\begin{equation}
\min_{\bm{P}^{\mathrm{nom}},\,\bm{p}_t}\;
\sum_{a}\,c^{\mathrm{ann}}_{a}\,P^{\mathrm{nom}}_{a}
\;+\;
\sum_{t}\,w_t \sum_{g}\, o_{g}\,p_{g,t},
\label{eq:pypsa_obj}
\end{equation}
subject to nodal balance, dispatch and storage constraints, and linearized power flow and line limits. Here $P^{\mathrm{nom}}_{a}$ and $c^{\mathrm{ann}}_{a}$ are the capacity and annualized cost of asset $a$; $p_{g,t}$ and $o_g$ are generator dispatch and marginal cost; and $w_t$ is the snapshot weight. LMP is the dual of nodal balance\cite{Bohn1984,Schweppe1988,Hogan1992}. Capital costs determine the expansion portfolio but are not added separately to the customer bill; customer wholesale exposure is measured through LMP.

\paragraph{Spatial resolution.}
The network contains 182 ERCOT buses identified by the \texttt{nerc\_reg} attribute. Topology, demand, and renewable availability follow the PyPSA-USA workflow\cite{Tehranchi2025}. We assign buses to West, North, South, or Houston using fixed geographic boundaries and calculate zonal LMPs as bus averages weighted by demand.

\paragraph{Temporal resolution and weather years.}
We solve 1,464 snapshots at three hour intervals from May~1 to October~30 (4,392 hours), covering the summer peak and the June--September four coincident peak (4CP) period. Experiments are repeated for weather years 2019--2023 using load and renewable availability for each year\cite{Staffell2018}. Unless noted, reported results for the announced 2030 scenario average these years. New capacity incurs its full annualized cost within the seasonal optimization; winter conditions are outside the study window.

\paragraph{Solver.}
Problems are solved with PyPSA~1.3.0\cite{PyPSA} and Gurobi~11.0.3\cite{Gurobi2024}, using the barrier method, linearized unit commitment, no transmission losses, and full dual assignment. Convergence and feasibility tolerances are $10^{-4}$ and $10^{-3}$.

\subsubsection*{Data center load scenarios}

Representative siting experiments place \SIlist{500;1000;3000;5000}{\mega\watt} at six development clusters: Tolar, Abilene, Childress, southern Dallas--Fort Worth, Austin, and Houston. These cases isolate the effect of where an otherwise identical load connects.

The announced 2030 scenario sums current, under construction, and planned capacity in 371 project records, geocodes each project, and aggregates it to the nearest model bus. The resulting scenario across 51 buses totals \SI{38.9}{\giga\watt}: \SI{18.5}{\giga\watt} in North, \SI{9.8}{\giga\watt} in South, \SI{8.4}{\giga\watt} in West, and \SI{2.3}{\giga\watt} in Houston. Each site runs at a capacity factor of 0.90, with a \(\pm5\%\) seasonal component and approximately 3\% noise, clipped to 80--100\% of nameplate. All announced capacity is supplied through the ERCOT network; supply behind the meter or on private grids is not represented. The derived capacity by bus is included in the code repository.

\subsubsection*{Generation, storage, and transmission expansion cases}

Candidate additions are combined-cycle (CCGT) and open-cycle (OCGT) gas turbines, utility solar, onshore wind, batteries with four hours of storage (no other duration is evaluated), and continuously expandable transmission corridors. Expansion of gas capacity is brownfield: it may expand only at buses where it exists initially. Candidate solar and wind units exist at every data center bus, whereas gas-fired generation is available at 58 buses (CCGT at 33 and OCGT at 46), including 26 of the 51 data center buses. Thus \SI{13.7}{\giga\watt} (35\%) of announced demand cannot receive dispatchable expansion at the same location. Technology costs follow PyPSA-USA and the NLR Annual Technology Baseline Moderate case\cite{Tehranchi2025,NREL2024}. Plant-level fuel costs follow PyPSA-USA; one gas unit in the 2021 base carried an implied fuel price below half the fleet median for that year, and its fuel component was reset to the fleet-median implied price before optimization. No other unit or year met this criterion.

A second candidate set adds one zero-capacity CCGT and one OCGT candidate at each of the 51 data-center buses. Each takes the annualized capital cost of its carrier, the fleet-median efficiency, variable cost and heat rate, the snapshot-wise fleet-median fuel cost, and full availability. Capital-cost premiums of 1.25, 1.5 and 2.0 on these candidates, uniform scaling of the announced load to 25, 50 and 75\%, the announced load without its already-operating share (30.9 of \SI{38.9}{\giga\watt}), and effective values of lost load of $10^{4}$ and $5\times10^{4}$~\$/MWh are run as 2023 sensitivities; the six representative sites are also run with gas candidates at the siting buses.

We vary the assets allowed to expand: none (\emph{dispatch}); generation only; generation and storage; generation and transmission; transmission only; or generation, storage, and transmission (\emph{full-tx}). The headline analysis uses \emph{full-tx}. Each case isolates one mechanism\cite{Brown2018,Victoria2020,Zeyen2023}.

\paragraph{Load shedding and the value of lost load.}
Feasibility is preserved by a generator representing load shedding with an effective value of lost load of \$5$\times$10\textsuperscript{6}/MWh. We recover unserved energy from its penalty cost and report MWh, share of demand, and intervals above the offer cap. Values below \SI{1}{\mega\watt\hour} per season are treated as solver tolerance. Cases with material shedding are not used as inputs for retail prices.

\paragraph{Offer cap and pricing rules.}
Prices paid by customers are truncated at ERCOT's \$5{,}000/MWh offer cap\cite{Potomac2024}, and values above the cap are reported separately as the above-cap component. In the planning model new capacity recovers its annualized cost through prices, much of it in these few intervals, so the truncated component excludes that part of the recovery. The share above the cap shifts with the gas price, the value of lost load, and the share of announced capacity that is built: in 2023 the increase before truncation stays at 24--27~\$/MWh across these cases, set by the cost of new capacity, while the truncated component varies as in Table~\ref{tab:ranges}. ERCOT itself keeps prices at or below \$5{,}000/MWh and recovers scarcity value through operating reserve demand curves\cite{Hogan2013,Potomac2024}. We represent this reserve-price rule by re-optimizing the coordinated expansion case and its baseline in each weather year with load shedding valued at \$5{,}000/MWh and an aggregate reserve demand curve, co-optimized with energy, whose parameters are fitted to ERCOT's posted curves; the network, candidate resources, season, and price estimator are unchanged. The wholesale component is then 2.84, 2.85, 2.89, 2.34, and 0.81~cents/kWh in 2019--2023. The 2023 value is lower because the reserve curve also raises the price of the 2023 baseline, in which existing demand comes closest to the capacity of the existing fleet; this uplift comes mostly from seven three-hour intervals. These runs shed up to \SI{15}{\giga\watt\hour} per season, less than 0.005\% of demand, and change reserve payments, which the main analysis holds constant, by at most 0.12~cents/kWh.

\subsubsection*{Wholesale price validation}

The model is linear with perfect foresight: it omits forecast errors, forced outages, startup bidding, and administrative scarcity adders, and its three-hour resolution smooths short price spikes\cite{Hogan2013,Potomac2024}.

\begin{figure}[!htbp]
\centering
\includegraphics[width=0.62\linewidth]{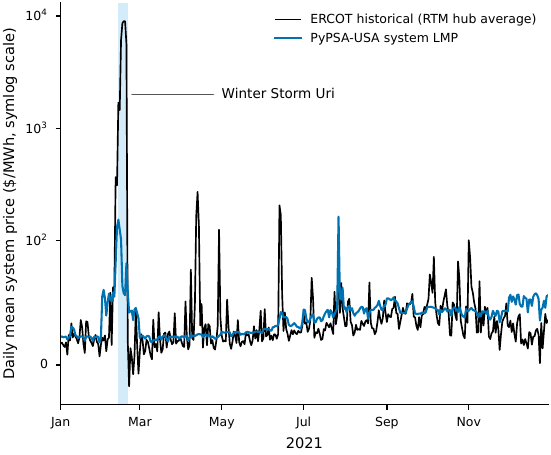}
\caption{\textbf{2021 validation of the PyPSA-USA ERCOT model against historical real-time prices.} Daily mean system prices (symlog scale). The simulation captures the timing of the Winter Storm Uri stress period (shaded) while smoothing the magnitude of scarcity pricing over short periods.}
\label{fig:validation}
\end{figure}

For May--October 2021, historical and modeled mean real-time prices are \$37.0 and \$37.5/MWh. Excluding Winter Storm Uri, the full year means are \$33.4 and \$36.9/MWh, and zonal means agree within 0--13\%. The model also reproduces the ordering across years of mean prices. It captures the timing but not the administered magnitude of Uri (Fig.~\ref{fig:validation})\cite{Busby2021,King2021}; daily correlations are 0.2--0.4 and the simulated 95th percentile is about 10\% lower outside Uri. Modeled prices therefore follow fundamentals and understate part of the short-period upper tail.

\subsection*{Transmission reinforcement and engineering cost model}

\subsubsection*{Texas2k network and contingency screening}

Analysis at the branch level uses case 1 of the 2025 Texas2k summer peak synthetic grid: 2,751 buses and 5,344 branches, comprising 3,993 lines at 115, 161, 230, and 500~kV and 1,351 transformers. The case serves \SI{85.8}{\giga\watt} of demand and contains approximately \SI{166}{\giga\watt} of generation\cite{Birchfield2017grid,Birchfield2017metric,Texas2k2025}. Bus and substation coordinates are supplied in the accompanying Texas2k geographic data; all screening uses the MATPOWER case. Its statistically generated topology reproduces the geographic footprint and aggregate characteristics of ERCOT without matching individual facilities, and results are reported as aggregate costs.

We use PYPOWER to solve DC power flows for the intact network (N-0) and for each branch outage that does not create an island (N-1). Each N-1 screen removes one operating branch, excluding radial elements with low capacity whose loss would island load, and records every monitored branch's worst loading across contingencies. The continuous thermal rating is applied in both the intact and contingency screens. Loadings between 80 and 100\% are retained as elevated utilization but do not trigger reinforcement, while loadings at or above 100\% do.

The attribution baseline is not free of violations: no branch is overloaded in the intact network, but N-1 screening under the worst contingency identifies 73 branches above their continuous ratings. We therefore distinguish violations newly caused by data center demand from those already present in the baseline.

\subsubsection*{Data center injection and upgrade costing}

We add 500, 1,000, 3,000, and 5,000~MW, plus announced capacity, at the same six regions used in the planning model. Loads operate at 0.98 power factor and are connected first to the regional bus with the highest voltage. Loads above \SI{3000}{\mega\watt}, or cases that fail to converge at one bus, are distributed across two to eight nearby buses. Tolar and Childress instead use all transmission buses within 30 and 75~km, respectively, with weights inversely proportional to distance favoring candidates at higher voltage levels. This represents the multiple points of interconnection of a gigawatt-scale cluster.

A violation is attributed to data center demand only when the corresponding branch is below 100\% in the baseline and reaches or exceeds 100\% after injection. Attribution uses loading in the intact network for N-0 and worst contingency loading for N-1.

Loading after injection determines the reinforcement. Branches at 100--110\% of rating are reconductored; those at 110--125\% receive high-temperature low-sag conductor; those at 125--150\% receive a parallel circuit; and loading above 150\% requires a new corridor or equivalent in the same voltage class. Line ratings are multiplied by 1.40, 1.75, 2.20, and 2.50 across these four tiers, with corresponding impedance reductions; transformer reinforcement factors range from 1.50 to 3.00. Line costs use voltage- and tier-specific MISO MTEP24 values, supplemented by PJM values for 500~kV facilities\cite{MISO2024,PJMM14B}. Transformer costs are assigned by voltage on the high side and reinforcement tier following ANSI/IEEE C57.12.00\cite{IEEEC57}. Route length is estimated as great circle distance multiplied by 1.30. Costs are reported in 2024 dollars as initial investment.

\subsubsection*{Coupling to the planning model and adequacy verification}

We evaluate two reinforcement strategies. The incremental strategy upgrades only violations newly attributable to data center injection, while the full remediation strategy also upgrades branches that already violate limits in the synthetic baseline. Rating and impedance changes are written back to the network and N-0 and N-1 screening is repeated once\cite{IEEE738,CIGRE763,NERCFAC008}. Residual and additional violations after this pass are reported without further reinforcement.

For the announced 2030 case in the 2023 weather year, the generation and storage added in the 2023 PyPSA coordinated expansion solution are mapped by location and technology onto nearby Texas2k buses before the full \SI{38.9}{\giga\watt} load is applied. Mapped additions replace rather than duplicate existing units. The attribution baseline contains the same expanded supply and original demand but no data centers. The coordinated-expansion portfolios of 2019--2022 and the 2023 portfolios with local gas candidates are mapped and screened in the same way, each against its own baseline; their charges are allocated on the 2023 coincident-peak base.

The \texttt{recoverNEW} case reinforces newly violating branches and supplies the primary engineering cost estimate; \texttt{recoverALL} additionally addresses baseline violations and provides a high-side sensitivity (\$35.59 against \$32.42~billion in the 2023 case). Expansion of an individual branch is capped at three times its initial rating, with larger requirements classified as structural residuals. The workflow omits reactive compensation, detailed routing of new corridors, permitting, and construction sequencing.

In the 2023 case, \texttt{recoverNEW} upgrades 519 target branches for N-1 violations; the single recovery pass clears 451 targeted violations, leaving 68 target violations and 143 violations across the updated grid in total.

\subsubsection*{Annualization of transmission cost}

We convert initial engineering investment to an annual revenue requirement using
\begin{equation}
\mathrm{CRF}=\frac{r(1+r)^{n}}{(1+r)^{n}-1},
\qquad
C^{\mathrm{ann}}_{\mathrm{TX}}
=\mathrm{CRF}\cdot C^{\mathrm{onetime}}_{\mathrm{TX}},
\label{eq:crf}
\end{equation}
with $r=8\%$ and $n=40$ years. The regulated transmission charge uses this annualized engineering cost.

\subsection*{Customer cost and retailer risk model}

\subsubsection*{Wholesale price adder}

The wholesale component of the effect on customer prices is the change in mean LMP between the baseline without data centers, solved on the same network with the same expansion permissions, and the coordinated expansion scenario, evaluated at the system level and for each zone,
\begin{equation}
A_{\mathrm{energy}}
=\overline{\lambda}^{\,\mathrm{full\_tx}}
-\overline{\lambda}^{\,\mathrm{baseline}} .
\label{eq:energy_adder}
\end{equation}
Both scenario and baseline prices are truncated at \$5{,}000/MWh before the adder paid by customers is calculated. The mean weights every snapshot equally and so gives the price change for a flat load; weighting snapshots by the demand of existing customers gives 1.53~cents/kWh in 2023 (Table~\ref{tab:ranges}).

\subsubsection*{Energy, congestion, and above-cap attribution}

To identify why the wholesale price changes, each bus price is separated into a system reference at each snapshot, capped locational congestion, and the amount above the offer cap:
\begin{equation}
\lambda_{i,t}
=\underbrace{\lambda^{\mathrm{ref}}_{t}}_{\text{energy}}
+\underbrace{\bigl(\min(\lambda_{i,t},\bar{\lambda})-\lambda^{\mathrm{ref}}_{t}\bigr)^{+}}_{\text{congestion}}
+\underbrace{\bigl(\lambda_{i,t}-\bar{\lambda}\bigr)^{+}}_{\text{above cap}},
\label{eq:decomp}
\end{equation}
where $\lambda^{\mathrm{ref}}_{t}=\min_i\min(\lambda_{i,t},\bar{\lambda})$ and $\bar{\lambda}=\$5{,}000/\mathrm{MWh}$. Averaging and differencing scenario and baseline values makes the billed wholesale adder the sum of the energy and congestion terms. The term above the cap is reported separately as the above-cap component and is not added to the customer bill. We do not apply this attribution where dual values from load shedding make the interpretation of the reference price invalid.

\subsubsection*{Regulated transmission charge}

We allocate annualized transmission cost using a method based on ERCOT's four coincident peaks\cite{Lyons2000,PUCT2023}. System 4CP is the mean coincident peak in June--September, computed from the three-hour snapshots,
\begin{equation}
\mathrm{4CP}_{\mathrm{sys}}
=\frac{1}{4}\sum_{m\in\{6,7,8,9\}}
\max_{t\in m}D_t ,
\label{eq:4cp}
\end{equation}
and a customer or load-serving entity (LSE) receives its share of the transmission cost of service (TCOS) according to its contribution to those four peaks:
\begin{equation}
s=\frac{D^{\mathrm{4CP}}_{\mathrm{LSE}}}
        {\mathrm{4CP}_{\mathrm{sys}}},
\qquad
\mathrm{TCOS}_{\mathrm{assigned}}
=s\,C^{\mathrm{ann}}_{\mathrm{TX}},
\qquad
A_{\mathrm{TCOS}}
=\frac{\mathrm{TCOS}_{\mathrm{assigned}}}{E_{\mathrm{LSE}}}.
\label{eq:tcos_adder}
\end{equation}
Equivalently,
\begin{equation}
A_{\mathrm{TCOS}}
=\frac{\tau\phi}{\SI{8760}{\hour}},
\qquad
\tau=\frac{C^{\mathrm{ann}}_{\mathrm{TX}}}
           {\mathrm{4CP}_{\mathrm{sys}}},
\end{equation}
where $\phi$ is the customer's ratio of 4CP demand to average load. We use $\phi=2.0$ for the representative residential customer and test $\phi=1.8$--2.5. We also test the allocation sensitivity in which data center demand is excluded from the 4CP denominator.

The residential effect on customer prices before any retail-provider (REP) risk premium is
\begin{equation}
A_{\mathrm{retail}}
=A_{\mathrm{energy}}+A_{\mathrm{TCOS}} .
\label{eq:retail_before_rep}
\end{equation}

\subsubsection*{Retail provider cash flow risk}

We simulate a stylized retail electricity provider with contracts at fixed prices (REP) following established risk models for load serving entities\cite{Bessembinder2002,Deng2006,Oum2006,Zhou2017,Mieth2021}. The REP serves a flat \SI{1000}{\mega\watt} load, hedges fraction $h$ at forward price $F$, and buys the remainder at capped spot price $\lambda_t$. The retail energy price $G$ is the EIA Texas residential price\cite{EIA2024} less the annual share for transmission and distribution. Procurement cost, margin, and cash are
\begin{align}
C_t &= hF+(1-h)\lambda_t, \\
m_t &= G-C_t-\mathrm{OpEx}-A_{\mathrm{TCOS}}, \\
\mathrm{cash}_t &= K+\sum_{\tau\le t}m_\tau L\Delta t,
\label{eq:cash_default}
\end{align}
and default occurs when cash becomes negative. The main simulation sets direct pass through of tariffs to zero, so the REP bears the transmission charge for a flat load while existing retail contracts remain fixed.

The baseline case using forward contracts uses the five-year mean of the baseline LMP, whereas the scenario case with repriced procurement uses the five-year mean of the scenario LMP. Each case uses 5,000 Monte Carlo paths per zone. A path first samples a weather year, then applies a block bootstrap with blocks of 48 hours to the LMP sequence in intervals of three hours and draws independent parameter shocks.

Base values are \$50~million starting capital, a hedge ratio of 0.70, a \$5/MWh forward premium, and \$3/MWh operating cost. Starting capital is lognormal with $\sigma=0.30$. The hedge ratio is truncated normal with $\sigma=0.10$ and support $[0,1]$. The forward premium is truncated normal with $\sigma=\$3$/MWh and support $[-20,50]$/MWh. Shocks to retail price and operating cost are independent zero-mean normal draws with standard deviations of 0.05 and 0.20, respectively. The LMP scale factor is lognormal with median 1 and $\sigma=0.10$. Draws from truncated normal distributions are clipped to their listed supports. The master random seed is 123; results using alternative random seeds differ by less than 0.2 percentage points at system level.

\subsubsection*{Sensitivity of customer pass through}

We define the required liquidity buffer as the cash needed to make the fifth percentile of minimum cash nonnegative,
\begin{equation}
B_{95}
=\max\!\bigl(0,\,-\,\mathrm{cash}^{\,p05}_{\min}\bigr).
\label{eq:buffer}
\end{equation}
Recovering this buffer over $Y$ years from annual load
$E=L\cdot\SI{8760}{\hour}$, with share passed through to customers $\rho$, gives
\begin{equation}
A_{\mathrm{REP}}
=\frac{B_{95}}{E\,Y}\rho,
\qquad
A_{\mathrm{total}}
=A_{\mathrm{energy}}+A_{\mathrm{TCOS}}+A_{\mathrm{REP}} .
\label{eq:rep_adder}
\end{equation}
We test $Y=1,3,5$ years and a deterministic grid of $\rho=0$--1. The REP risk term is kept separate from the residential wholesale and regulated transmission effect because the two arise at different stages of the retail contract cycle.

\subsubsection*{Channel separation and model coupling}

Wholesale prices, regulated transmission cost, and REP risk are kept separate throughout. PyPSA capital costs select the planning solution and are not billed separately. The planning model prices electricity on a 182-bus network with corridor-level transmission and has no individual branches, contingencies, or facility costs, whereas the Texas2k screening costs specific reinforcements under the N-1 criterion used to identify reinforcement needs, but does not price energy. Their transmission costs also differ: in 2023 the coordinated expansion adds corridor capacity costing \$0.14~billion per year at PyPSA-USA cost assumptions, against \$2.719~billion per year for the engineering reinforcement. The PyPSA and Texas2k engineering models are coupled in sequence: wholesale prices are calculated on the optimized planning network, while the regulated charge is based on engineering reinforcement at the branch level. We do not re-estimate the LMP that would result from installing that exact engineering portfolio. The wholesale and regulated components are estimated separately and summed, without calibration to observed bills.

Distribution charges, ancillary services, taxes, fixed charges, and existing retail margins are held constant. The customer cost estimate is an incremental, system-level quantity.

\section*{Data availability}
The Texas2k synthetic network is publicly available from the Electric Grid Test Case Repository at Texas A\&M University\cite{Texas2k2025}. PyPSA-USA input data are available through the open PyPSA-USA workflow\cite{Tehranchi2025}. Historical ERCOT prices are available from ERCOT\cite{ERCOT2024} and retail prices from the U.S. Energy Information Administration\cite{EIA2024}. Processed simulation outputs supporting every figure in this paper, including the capped locational marginal price adders, the attribution of the wholesale price adder, results for unserved energy for both the announced 2030 and representative siting experiments, contingency results from stepped injection, tables of upgrade costs, and the retail provider Monte Carlo summaries, are available in the code repository listed under Code availability, together with the scenario definitions and the derived mapping of data-center capacity to model buses.

\section*{Code availability}
The scenario definitions, PyPSA-USA configurations, wholesale-price and customer-cost analyses, retail-provider Monte Carlo model, and figure-generation scripts are available at \url{https://github.com/LDZhou/texas-datacenter-electricity-costs} under the MIT license. The \texttt{texas2k/} directory contains the PYPOWER DC contingency-screening and transmission-reinforcement workflow, together with its input tables and reference outputs. The repository documents the steps linking capacity expansion, engineering reinforcement, and residential electricity costs.

\FloatBarrier

\section*{Acknowledgements}
This work was authored in part by the National Laboratory of the Rockies (NLR) for the U.S. Department of Energy (DOE), operated under Contract No. DE-AC36-08GO28308. Funding for NLR authors only provided by U.S. Department of Energy Office of Electricity. The views expressed in the article do not necessarily represent the views of the DOE or the U.S. Government. The U.S. Government retains and the publisher, by accepting the article for publication, acknowledges that the U.S. Government retains a nonexclusive, paid-up, irrevocable, worldwide license to publish or reproduce the published form of this work, or allow others to do so, for U.S. Government purposes.

\section*{Author contributions}

Conceptualization, X.Z. and F.D.;
methodology, L.Z., W.X., X.Z., and F.D.;
investigation, L.Z., W.X., X.Z., and F.D.;
writing--original draft, L.Z. and W.X.;
writing--review \& editing, L.Z., W.X., J.W., X.Z., and F.D.;
funding acquisition, F.D.;
resources, F.D. and J.W.;
supervision, X.Z. and F.D.

\section*{Competing interests}
The authors declare no competing interests.

\end{document}